\documentclass[graybox]{svmult}

\usepackage{mathptmx}       
\usepackage[scaled=.5]{helvet}         
\usepackage{courier}        
\usepackage{type1cm}        
\usepackage{makeidx}         
\usepackage{graphicx}        
\usepackage{multicol}        
\usepackage[bottom]{footmisc}
\usepackage[figuresleft]{rotating}
\usepackage{hyperref}

\makeindex             

\begin{document}

\title*{Search for cosmological $\mu$ variation from high-redshift H$_2$ absorption; a status report}
\titlerunning{Search for cosmological $\mu$ variation from high-redshift H$_2$ absorption}
\author{W. Ubachs, J. Bagdonaite, M.T. Murphy, R. Buning, and L. Kaper}
\institute{Wim Ubachs \at 1) Institute for Lasers, Life and Biophotonics, VU University Amsterdam, de Boelelaan 1081, 1081 HV Amsterdam, The Netherlands; \email{wimu@nat.vu.nl}
\and Julija Bagdonaite \at Department of Astronomy, Vilnius University, Vilnius,
Lithuania; Also at 1); \email{j.bagdonaite@vu.nl}
\and Michael T. Murphy \at Centre for Astrophysics and Supercomputing,
Swinburne University of Technology, Melbourne, Victoria 3122, Australia; \email{mmurphy@swin.edu.au}
\and Ruth Buning \at At 1); Present address: Leiden Institute of Physics, Leiden University, The Netherlands; \email{buning@physics.leidenuniv.nl}
\and Lex Kaper \at Astronomical Institute Anton Pannekoek, Universiteit van Amsterdam,
1098 SJ Amsterdam, The Netherlands; Also at 1); \email{l.kaper@uva.nl}
}
\maketitle

\abstract*{Fill in the same text here.}

\abstract{Observations of H$_2$ spectra in the line-of-sight of distant quasars may reveal 
a variation of the proton-electron mass ratio $\mu=m_p/m_e$ at high redshift, typically for $z>2$. Currently four high-quality systems
 (Q$0347-383$, Q$0405-443$, Q$0528-250$ and J$2123-005$) have been analyzed returning a constraint
 $\Delta\mu/\mu < 1 \times 10^{-5}$. We present data and a $\mu$-variation analysis of another system, Q$2348-011$ at redshift $z_{abs}=2.42$, delivering $\Delta\mu/\mu = (-1.5 \pm 1.6) \times 10^{-5}$.
 In addition to observational data the status of the laboratory measurements is reviewed. The future possibilities of deriving a competitive constraint on $\Delta\mu/\mu$ from the known high-redshift H$_2$ absorbers is investigated, resulting in the identification of a number of potentially useful systems for detecting $\mu$-variation.
}

\section{Cosmological $\mu$ variation and the laboratory database}
\label{sec:1}
The proton-electron mass ratio $\mu=m_p/m_e$ is one of the constants suitable for targeting a possible variation of a physical constant on a cosmological time scale.
Although $\mu$ is a dimensionless constant, lending itself for an observational approach, it is not a \emph{fundamental} constant in a strict sense, like the fine-structure constant $\alpha$, because $\mu$ involves all the binding physics of the quarks inside the proton. However, its close connection to $m_e/\Lambda_{QCD}$ makes it the parameter determining the measure of the strong force with respect to the electroweak sector, and is therewith of fundamental importance.
Variation of $\mu$ can be straightforwardly derived from a comparison between spectral lines observed at vastly differing redshifts.
Values of wavelengths at high redshift $\lambda_z^i$ are compared to laboratory wavelengths $\lambda_0^i$ for as many as possible H$_2$ lines~\cite{Thompson1975} using sensitivity coefficients for each individual spectral line $K_i=d \ln \lambda_i/ d \ln \mu$. These $K_i$ are calculable to the 1\% accuracy level~\cite{Ubachs2007}.

The laboratory wavelengths of the spectral lines in the Lyman and Werner bands of H$_2$ have been investigated at ever improved accuracy using tunable vacuum ultraviolet (VUV) lasers since over a decade~\cite{Hinnen1994,Philip2004,Ubachs2004,Hollenstein2006,Reinhold2006,Ivanov2008a}, also including spectra of HD~\cite{Hinnen1995,Ivanov2008b}. In addition the novel Fourier-transform spectrometer in the VUV range at the Soleil synchrotron was used to perform direct absorption spectra focusing on HD to cover the entire range of Lyman and Werner bands for this deuterated molecule~\cite{Ivanov2010}. The most accurate wavelength positions are obtained from a third indirect spectroscopic method based on the highly accurate determination of level energies for certain anchor levels~\cite{Hannemann2006}, where the determination of Lyman and Werner wavelengths results from an independent measurement of spacings between excited states~\cite{Salumbides2008}. The full set of recommended wavelengths is reported in Bailly \emph{et al.}~\cite{Bailly2010}. Another compilation of these data, both for H$_2$ and HD is given in the appendix of Malec \emph{et al.}~\cite{Malec2010}, also including a listing of $K_i$ sensitivity coefficients~\cite{Ubachs2007,Ivanov2010}, oscillator strengths of the absorption lines~\cite{Abgrall1994}, and of damping coefficients~\cite{Abgrall2000}.

A quasar absorption spectrum can be treated in different ways to detect a possible $\Delta\mu/\mu$. A straightforward method is to derive line positions by fitting H$_2$ resonances, and thereupon compare the results with the laboratory wavelengths; by deriving reduced redshifts $\zeta_i=(z_i-z_A)/(1+z_A)$ with $z_i$ the redshift determined for each line ($z_i=(\lambda_z^i/\lambda_z^0)-1$) and $z_A$ the redshift of the absorber, such a method gives a graphical insight into a possible variation of $\mu$~\cite{Ubachs2007,Reinhold2006}. In contrast to this line-by-line fitting method a \emph{comprehensive} fitting method has advantages. There, an H$_2$ spectrum is generated, based on the available molecular parameters (wavelengths~\cite{Bailly2010,Malec2010}, oscillator strengths~\cite{Abgrall1994}, and damping parameters~\cite{Abgrall2000}). The value for $\Delta\mu/\mu$ is then a single fitting parameter comparing the observed spectrum with a model spectrum; this is accomplished via least squares fitting with the {\sc VPFIT} software~\cite{VPFIT}. Each line is convolved with the instrumental profile (pertaining to the spectral resolving power $R = \lambda/\Delta\lambda$ of the spectrometer), a natural linewidth $\Gamma$, and a Doppler parameter $b$. The main advantage of the method is that a composite velocity structure can be addressed. The column densities $N(J)$ for each rotational state can be tied between velocity components or left independent. Further details and advantages of the comprehensive fitting method are discussed in Refs.~\cite{Malec2010,King2008}.
By this means the spectral information on overlapping lines and of saturated lines can be included. In the example of the analysis of the Keck-HIRES spectrum of the quasar J$2123-005$ it is shown how the velocity structure can be investigated in much detail in terms of computed \emph{composite residuals} which demonstrate a complex velocity structure that should be accounted for~\cite{Malec2010}. These optimized comprehensive fitting methods were also implemented in the analyses of the three systems Q$0347-383$, Q$0405-443$ and Q$0528-250$ by King \emph{et al.}~\cite{King2008}.

\begin{center}
\begin{figure}[t]
%
\includegraphics[scale=0.5]{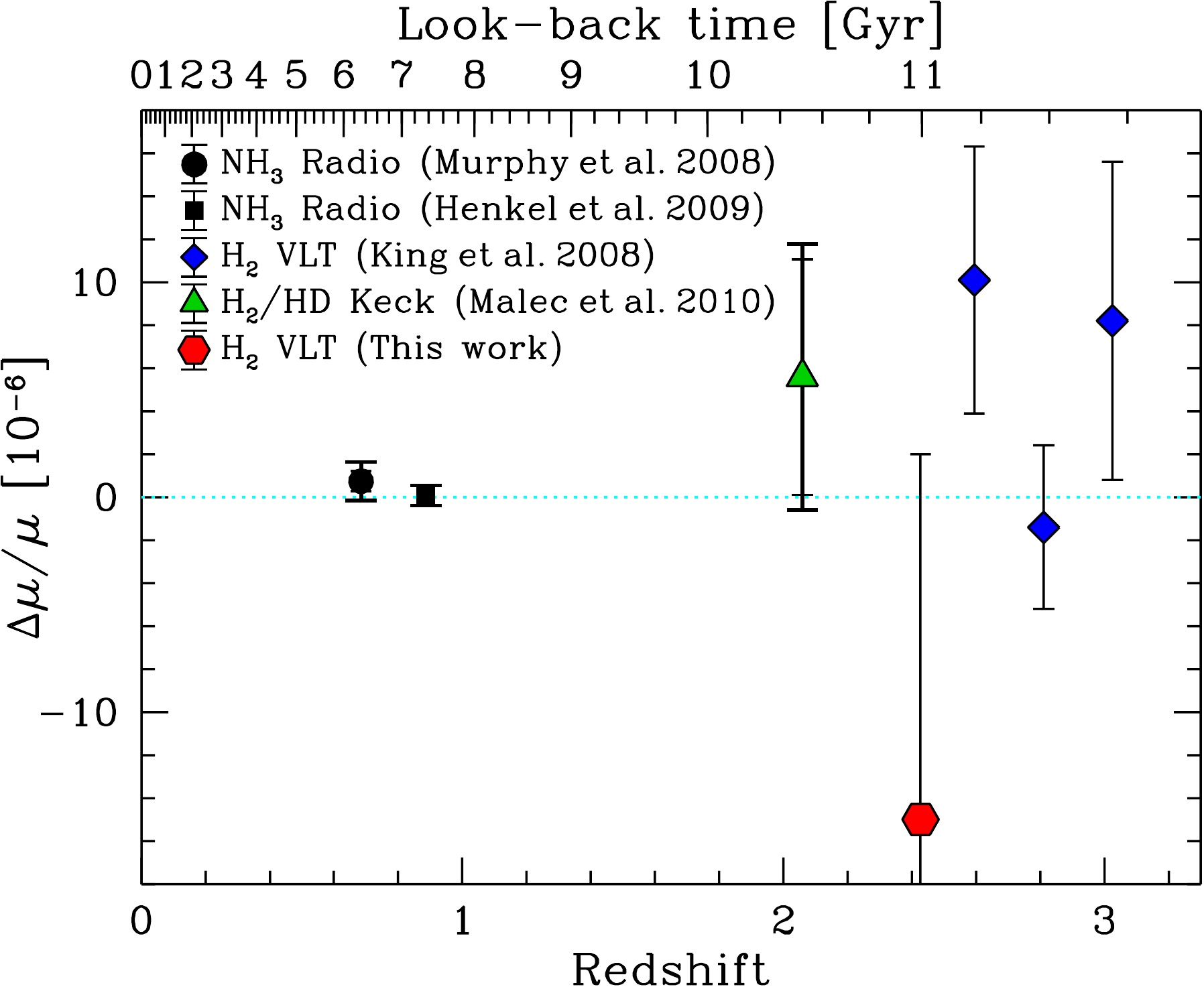}
\caption{Current result on cosmological $\mu$-variation based on the five available high-quality H$_2$ absorption systems as a function of redshift:
blue points relate to Q$0347-383$, Q$0405-443$ and Q$0528-250$ \cite{King2008}, green is the result from J$2123-005$ \cite{Malec2010}, while the black points relate to the ammonia data \cite{Murphy2008,Henkel2009} for redshifts $z<1$. The red hexagonal point represents the result of the present analysis of Q$2348-011$.
}
\label{fig:1}       
\end{figure}
\end{center}

The resulting constraints on $\Delta\mu/\mu$ for the four systems mentioned are shown in Fig.~\ref{fig:1}. The figure also inculdes the result of the Q$2348-011$ analysis, details of which will be presented in the next section. The tighter constraints stemming from radio-astronomical observations of NH$_3$ at redshifts $z<1$ are shown as well~\cite{Murphy2008,Henkel2009}. The data show the present status of a possible variation of the proton-electron mass ratio to be $\Delta\mu/\mu < 1 \times 10^{-5}$ at redshifts $z=2- 3.5$.

\section{VLT-UVES observation of Q2348$-$011}
\label{sec:2}

H$_2$ absorption features associated with the Damped Lyman-$\alpha$ (DLA) system at $z=2.42$ along the sightline of Q$2348-011$ (SDSS J$235057.88-005209.8$, $z_{em} \simeq 3.02$) were first reported by Petitjean \emph{et al.}~\cite{Petitjean2006}. Since then this system has been the subject of several investigations but none of them focused on $\mu$-variation (e.g. Ledoux \emph{et al.}~\cite{Ledoux2006}, Noterdaeme \emph{et al.}~\cite{Noterdaeme2007}), but rather on the molecular column densities, the metallicity and the physical conditions of the gas contained in this DLA-system. For an investigation of $\mu$-variation new observations of well-calibrated spectra of Q$2348-011$ were recorded, and will be presented here.

\begin{figure}
\includegraphics[scale=0.63]{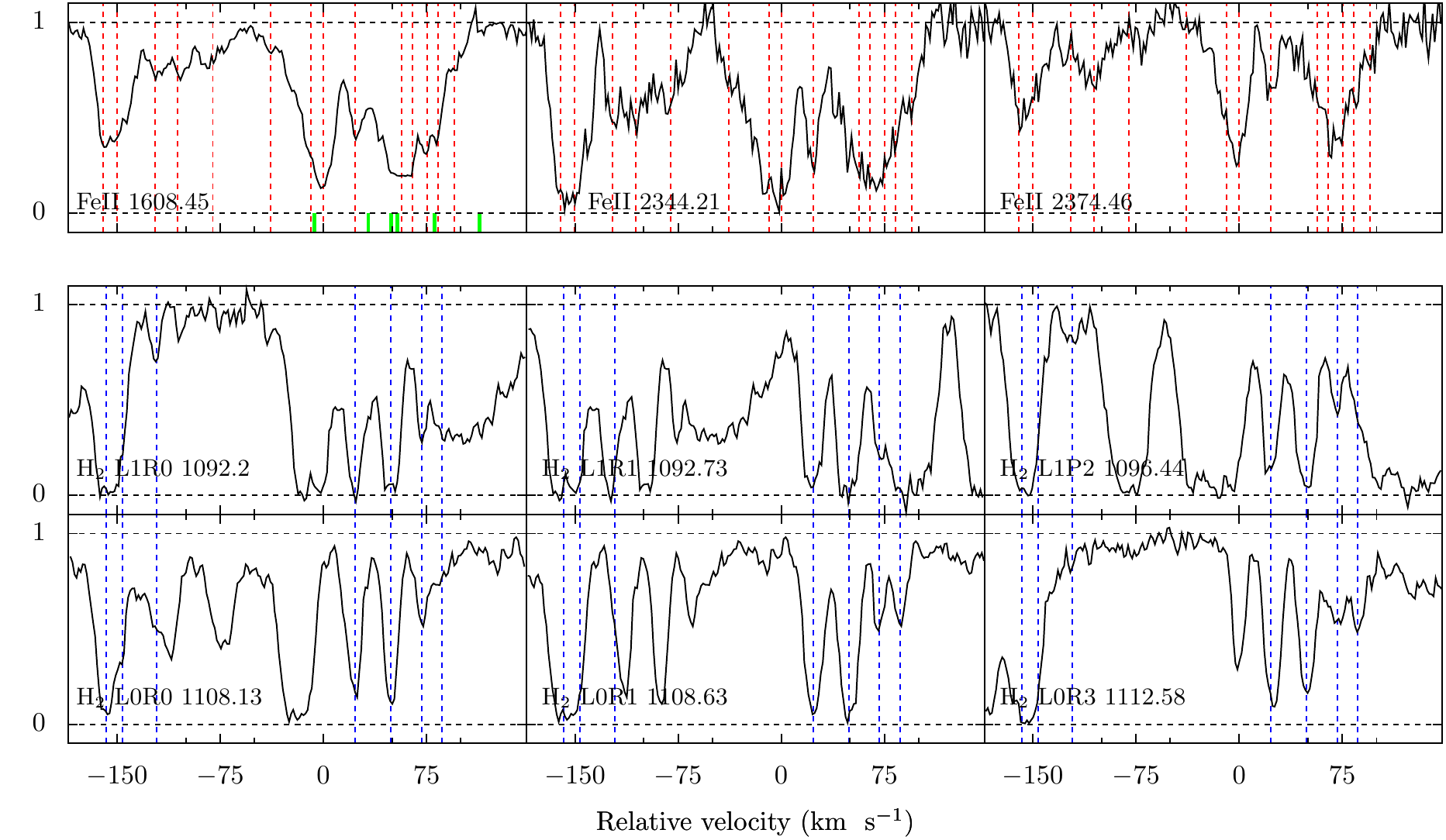}
\caption{Lower panel: H$_2$ velocity structure in the Q$2348-011$ DLA system at $z=2.42$; upper panel: Fe II velocity structure of the same absorber. The positions of 7 different velocity features of H$_2$ are indicated by the blue dashed lines, while the positions of 14 velocity components of Fe II are indicated by the red dashed lines. Note that the strongest velocity component in Fe II absorption profile (at 0 km s$^{-1}$) does not have an associated H$_2$ component. The cleanest Fe II transition (1608.45 \AA) is blended with Si IV (1402.77 \AA) of an additional absorber at $z=2.93$. The green ticks in the Fe II panel indicate positions of the overlapping Si IV lines.
}
\label{fig:2}
\end{figure}

\subsection{Data}
The absorption spectrum of the quasar Q$2348-011$ was obtained with the Ultraviolet and Visual Echelle Spectrograph (UVES) at the Very Large Telescope (VLT) of the European Southern Observatory (ESO) in Paranal, Chile. Observations were carried out in visitor mode on four consecutive nights (2007 August 18-21). Q$2348-011$ is not a particularly bright object ($R_{mag} = 18.31$) so long exposures were needed. An overall of 15 spectra were recorded with an exposure time between 3600 and 4800 seconds each; they make up for a combined total of 19.25 hours of observation. The slit widths were set to $0.''8 - 1.''0$ and $0.''7 - 0.''8$ in, respectively, blue and red ranges. They were kept constant for all exposures irrespective of the noticeably varying seeing conditions, which were on average $1.''2$.
These settings yield a resolution of $R \sim 51\,700$ (in the blue-UV range where all H$_2$ lines are observed) and $55\,300$ (in the red). This corresponds to a resolution element of 2.46 and 2.31 km s$^{-1}$ in terms of velocity units, respectively. For optimum wavelength calibration the 15 science exposures were each attached with a ThAr calibration frame, obtained immediately after each science exposure. Special care was taken to use pre-selected Th-Ar reference lines following optimization procedures~\cite{Murphy2007}.
The final spectrum of Q$2348-011$ covers (vacuum-heliocentric) wavelengths 3572-9467 $\textrm\AA$, with gaps at 4520-4621 and 7505-7665 $\textrm\AA$. Parts of the spectrum are displayed in Fig.~\ref{fig:2}.

\subsection{Analysis}

\begin{figure}
\includegraphics[scale=0.85]{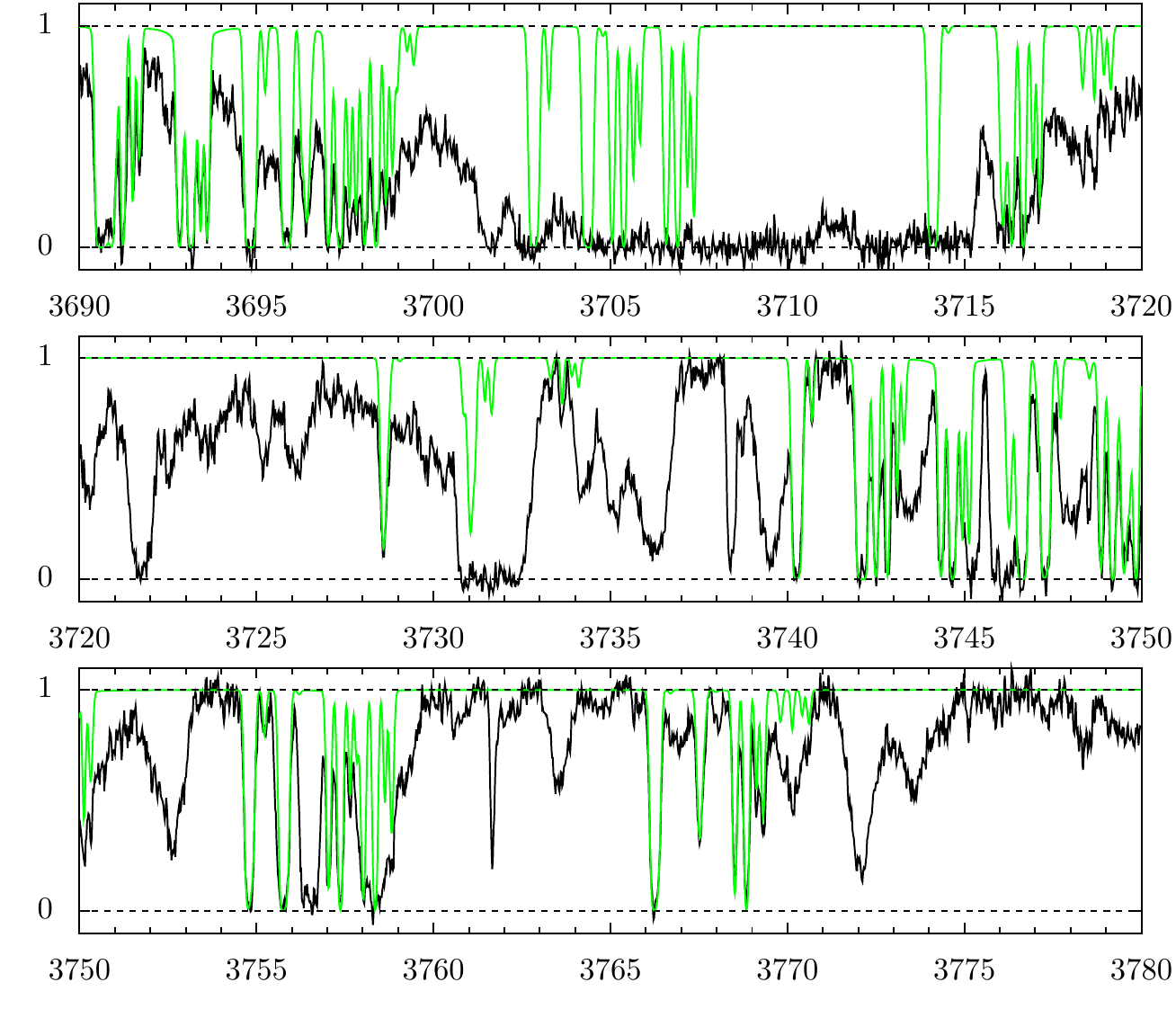}
\caption{H$_2$ absorption model superimposed on a portion of the Q$2348-011$ spectrum. The strong feature centered at 3710 \AA \ is Lyman-$\beta$ of the DLA at $z = 2.62$.
}
\label{fig:3}
\end{figure}

The wavelength range covered by the VLT-UVES spectrum provides 58 H$_2$ transitions for rotational levels $J=0-5$. The H$_2$ features fall between 3580 and 3860 $\textrm\AA$ for this absorber at $z=2.42$. This part of the combined spectrum has a signal-to-noise ratio (S/N) of 25. Note that no associated HD spectral features have been detected at a significant level. All the transitions observed arise from the Lyman band. Molecular hydrogen is present in 7 velocity features (the two left-most are blended together but they count at least as two), so in principle this spectrum would provide a sample of \(58\times7\) absorption lines. However, besides the DLA at $z=2.42$ (neutral content $(\log N($H I$) = 20.50\pm0.10)$, \cite{Ledoux2006}), which contains molecular hydrogen, there is an additional DLA at $z=2.62$, which has a larger column density of neutral hydrogen $(\log N($H I$) = 21.30\pm0.08)$ \cite{Noterdaeme2007}), but no signs of molecular hydrogen.
The strong Lyman-$\beta$ feature of the second DLA ($z=2.62$) at $3710$ \AA \ obscures 4 H$_2$ transitions (24 velocity components) and damps some others (see Fig.~\ref{fig:3}), while also the Lyman-$\gamma$ blends 4 H$_2$ lines (28 velocity components). This DLA gives rise to a cutoff in the absorption spectrum at $3580$ \AA, which makes this DLA an unfortunate coincidence along the sightline of Q$2348-011$.
In addition there exist strong H I absorbers, almost at the column densities of sub-DLA's, at $z=2.73$ and $z=2.93$ obscuring several H$_2$ lines.

Another distinctive feature of this spectrum is self-blending of H$_2$: since the velocity structure of H$_2$ consists of numerous components spread over \(\sim\) 300 km s$^{-1}$, nearby lying lines exhibit mutual overlap of velocity components. For example, velocity components 4-7 of H$_2$ L0R1 lie in the same region as velocity components 1-3 of H$_2$  L1P5. This kind of overlapping can be disentangled using the comprehensive fitting method, if a sufficient number of non-overlapping lines is available. However, the non-overlapping lines are blended in addition with the absorbers of the Lyman-\(\alpha\) forest, and the overall amount of clear features is strongly reduced.

Some of the molecular hydrogen features are contaminated by metal lines associated with the two DLAs, namely, L2R0 with Si II (1020.69 \AA) at $z=2.62$, and L3R1, L3P1, L1P2, L1R3, L0P2, L0R3 are blended  with various Fe II lines associated with the $z=2.42$ DLA. Further, L1R5 and L1P4 may be blended with an additional C IV system at $z=1.44$ but this is difficult to reconstruct unambiguously. Generally, the sightline towards the quasar Q$2348-011$ is rich in interesting absorbers: besides the two DLAs there are other strong metal absorbers at $z=2.97$, $2.93$, $2.72$, $2.58$ (C IV, Si IV, N V), and at $z=0.86$, $0.77$ (Mg II). The latter is remarkable in having numerous components spread over 750 km s$^{-1}$. Noteworthy coincidence is that the cleanest Fe II transition (1608.45 \AA) associated with $z=2.42$ is blended with Si IV (1402.77 \AA) of the absorber at $z=2.93$ (see Fig.~\ref{fig:2}). Also note that the most prominent velocity component in metal transitions does not have an associated H$_2$ component.

For the final fit deriving $\Delta\mu/\mu$ 35 regions with 36 different molecular hydrogen transitions have been selected. Each line is modeled as a Voigt profile. In the comprehensive fitting method the parameters (column density $N$, Doppler parameter $b$, and position $z$) are connected for different lines. Each $J$-level for H$_2$ has a different ground state population, so all transitions from the same $J$-level have the same value for $N$, $b$, and $z$. It is also assumed that a given velocity component has the same $z$ and $b$ value in all $J$-levels; physically this can be interpreted that the molecular hydrogen of all states resides in the same absorbing cloud. The value for $\Delta\mu/\mu$ is a single fitting parameter for all molecular hydrogen transitions. The absorption model is optimized to fit data by using the {\sc VPFIT} software.

As analyzed by Malec~\emph{et al.} in the example of the J$2123-005$ quasar system, more than one velocity component can underlie a molecular absorption feature as distinguishable by inspection. Their presence can be confirmed or rejected relying on goodness-of-fit and by inspecting \emph{composite residuals}~\cite{Malec2010}. As the two left-most features of H$_2$ absorption are blended, more components could in principle be present close to or between the two blended features. A fit with 7+1 velocity components (VC) instead of 7, where the additional VC was included between the two left-most VCs, has improved the $\chi^2_{\nu}$ per degree-of-freedom by several hundredths, but the additional velocity component was rejected as unnecessary in transitions from $J=4$ and $J=5$ rotational levels. The 7+1 VC model is regarded as the fiducial result, because it gives the best match with the observed spectrum, i.e. it is statistically preferred as it returns the smallest $\chi^2_{\nu}$. A more detailed analysis will be presented in an upcoming paper \cite{JB2011}.

\section{Results}

\begin{sidewaystable}
    \caption{Parameters and 1-$\sigma$ statistical uncertainties pertaining to the 7+1 VC fiducial fit model for the molecular hydrogen absorption in the Q2348$-$011 spectrum. The 8 velocity components are at redshifts $z$ and have Doppler widths $b$. The second row gives the velocity relative to $z =$ 2.426318 in units of km s$^{-1}$.}
    \centering
    \begin{tabular}{c c c c c c c c c}
    \hline\noalign{\smallskip}
    VC & $z$ & $b$ & \multicolumn{6}{c}{$\log N(\mathrm{H_2})\,[\mathrm{cm^{-2}}]$} \\
       & $\Delta v$ [km s$^{-1}$] & [km s$^{-1}$] & J=0 & J=1 & J=2 & J=3 & J=4 & J=5  \\ [0.5ex]
    \noalign{\smallskip}\svhline\noalign{\smallskip}
    1a & 2.424509(1) & 6.73 $\pm$ 0.15 & 15.27 $\pm$ 0.15 & 16.45 $\pm$ 0.05 & 15.50 $\pm$  0.14 & 15.92 $\pm$  0.07 & 15.30 $\pm$  0.03 & 15.12 $\pm$  0.02\\
       & -158 & & & & & & & \\
    1b & 2.424546(4) & 9.94 $\pm$ 0.29 & 15.45 $\pm$ 0.08 & 15.53 $\pm$ 0.40 & 15.79 $\pm$ 0.07 & 15.67 $\pm$  0.10 & -- & --  \\
       & -155 & & & & & & & \\
    2 & 2.424652(2) & 1.25 $\pm$ 0.24 & 14.97 $\pm$ 0.12 & 16.98 $\pm$ 0.50 & 14.37 $\pm$  0.27 & 14.77 $\pm$  0.21 & 14.20 $\pm$  0.06 & 13.62 $\pm$  0.13 \\
       & -146 & & & & & & & \\
    3 & 2.424929(2) & 4.15 $\pm$ 0.33 & 14.29 $\pm$ 0.03 & 14.47 $\pm$ 0.05 & 14.51 $\pm$  0.03 & 13.67 $\pm$ 0.34 & 12.95 $\pm$ 1.00 & -- \\
       & -122 & & & & & & & \\
    4 & 2.426582(1)  & 4.56 $\pm$ 0.13 & 15.66 $\pm$ 0.02 & 15.84 $\pm$ 0.10 & 15.60 $\pm$  0.03 & 15.62 $\pm$  0.04 & 14.29 $\pm$ 0.04 & 13.81 $\pm$  0.11 \\
       & +23 & & & & & & & \\
    5 & 2.426883(1)  & 2.97 $\pm$ 0.08 & 16.24 $\pm$ 0.08 & 17.98 $\pm$ 0.04 & 17.27 $\pm$  0.10 & 16.99 $\pm$  0.13 & 14.31 $\pm$ 0.08 & 14.08 $\pm$  0.07 \\
       & +50 & & & & & & & \\
    6 & 2.427136(1)  & 3.10 $\pm$ 0.12 & 14.72 $\pm$ 0.05 & 15.22 $\pm$ 0.03 & 14.93 $\pm$  0.04 & 14.86 $\pm$  0.06 & 14.09 $\pm$ 0.06 & 13.67 $\pm$  0.16 \\
       & +72 & & & & & & & \\
    7 & 2.427310(1)  & 4.36 $\pm$ 0.22 & 14.45 $\pm$ 0.05 & 15.27 $\pm$ 0.03 & 14.75 $\pm$  0.05 & 15.09 $\pm$  0.05 & 14.29 $\pm$ 0.04 & 13.89 $\pm$  0.16\\
       & +87 & & & & & & & \\ [0.5ex]
    \hline
    \end{tabular}
    \label{table:results}
\end{sidewaystable}

The fiducial absorption model led to the following result: $(\Delta\mu/\mu)_{7+1VC} = (-1.5 \pm 1.6) \times 10^{-5}$.
Table \ref{table:results} provides molecular cloud properties with their statistical uncertainties as derived from the final fitting attempt when $\Delta\mu/\mu$ was set as a free parameter.

The analysis of H$_2$ absorption in the Q2348$-$011 spectrum yields a competitive result on $\Delta\mu/\mu$, but it is three times less tight compared to the constraints from previous analyses (see Fig.~\ref{fig:1}). One of the major causes that led to the relatively large uncertainty is the low S/N of the spectrum. The second cause is that the spectrum is obscured by the neutral hydrogen features of the additional strong DLA at $z=2.62$. The H$_2$ transitions falling at the shortest wavelengths covered by the spectrum are especially useful in $\mu$-variation analysis, since they exhibit the larger $K$-coefficients (transitions in the Lyman band with higher vibration excitation). In the spectrum of Q$2348-011$ they are not detected due to the H I absorption produced by the additional DLA.
Although the total number of H$_2$ transitions available in this spectrum is reduced, it is at least partially compensated by the presence of numerous velocity components associated with each transition. Altogether, each transition contributes a higher information content when it is imprinted in several velocity components. The complex absorption structure was successfully modelled by means of the comprehensive fitting method.

\section{Potential high-redshift H$_2$ absorbers for future detection of $\mu$-variation}
\label{sec:3}

Detection of cosmological $\mu$-variation, or putting strong constraints on the variation, should preferably be based on large numbers of high redshift H$_2$ absorbing systems. For comparison the recent study on a possible spatial dipole for the fine structure constant $\alpha$ relies on the analysis of almost 300 absorbing systems~\cite{Webb2010}. However, for H$_2$, up to now only five systems (Q$0347-383$, Q$0405-443$, Q$0528-250$, J$2123-050$ and Q$2348-011$) of sufficient quality are found, and the results obtained in comprehensive fitting procedures are summarized in Fig.~\ref{fig:1}. We note that alternative line-by-line fitting procedures on the Q$0347-383$ and Q$0405-443$ systems have recently been reported~\cite{Wendt2008,Thompson2009,Wendt2010}. Taking the weighted mean of the H$_2$ results of Fig.~\ref{fig:1} yields $\Delta\mu/\mu = (2.9 \pm 2.6) \times 10^{-6}$ for redshift ranges between $z=2.0 - 3.1$, which may be considered as the status by September 2010. The data provide firm evidence that $\mu$-variation is below the $10^{-5}$ level for look-back times in the range of 10-12 Gyrs.

\begin{figure}
\includegraphics[scale=0.35]{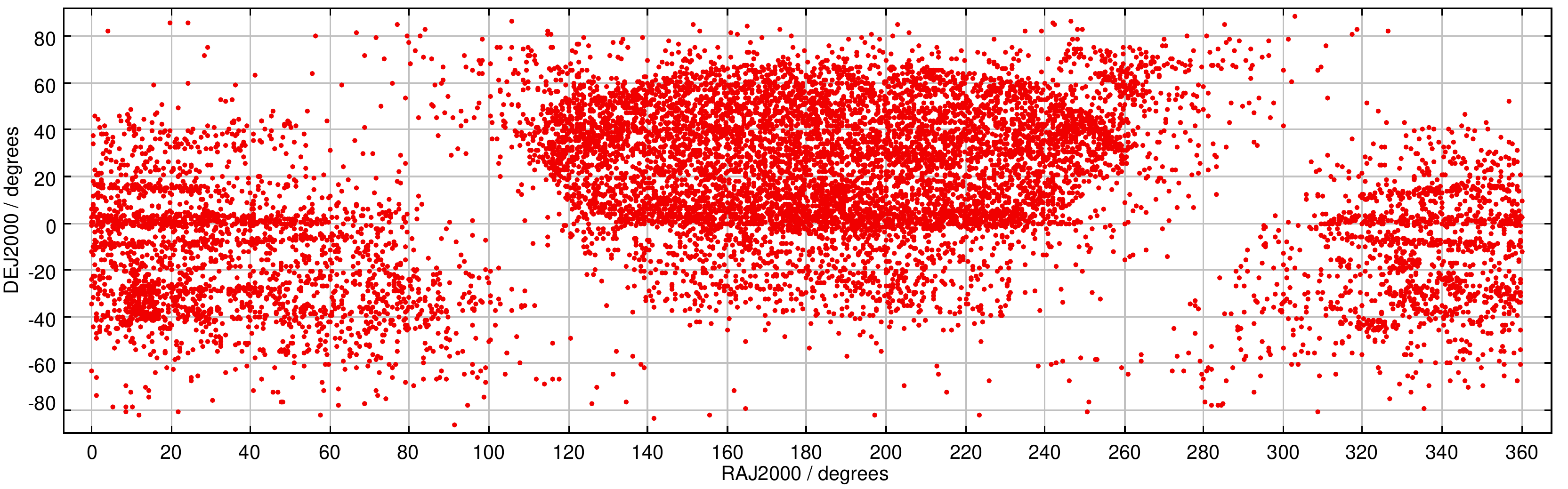}
\caption{Known quasars with magnitude $V<18$ as obtained from~\cite{Veron2010}.
}
\label{fig:4}
\end{figure}

A recent analysis of the J$1337+315$ spectrum by Srianand \emph{et al.}~\cite{Srianand2010} yielded a constraint only at the $10^{-4}$ level. This is due to the combination of low brightness of the quasar ($R_{mag} = 18.08$) and low H$_2$ content at the absorbing cloud ($\log_{10}N$(H$_2$)$ = 14.1$). The present analysis of the Q$2348-011$ H$_2$ absorption system demonstrates, that even when the S/N of a spectrum is low, a competitive constraint on $\Delta\mu/\mu$ can still be delivered if other circumstances are favorable. In case of Q$2348-011$ the low S/N is compensated by the presence of at least 7 sufficiently strong velocity components. However, in this regard the Q$2348-011$ absorbing system is rather an exception than a rule. In other known DLAs H$_{2}$ absorption is usually detected in no more than three visually distinguishable spectral features.
 
This raises the question how many of the thousands of quasar systems on the sky (see Fig.~\ref{fig:4} and~\cite{Veron2010}), of which there are some 1000 identified as DLAs could produce a competitive result, i.e. at the level $\delta(\Delta\mu/\mu) \leq 5 \times 10^{-6}$. Based on our experience in analyzing spectra, we postulate that high quality H$_2$ absorption systems should obey the following conditions:
\begin{itemize}
\item{An flux of $V_{mag}< 18$ of the quasar background source for providing a S/N of about 50 after a reasonable amount of data collection (15-20 h).}
    \end{itemize}
\begin{itemize}
\item{A column density of $N$(H$_2$) preferably in the range between $10^{14}$ and $10^{18}$ cm$^{-2}$ yielding sufficient but not too saturated H$_2$ absorption.}
    \end{itemize}
\begin{itemize}
\item{An H$_2$ absorption system at redshift $z>2$; only then a sufficient number of lines will shift into the atmospherically transparent window and the throughput window of Keck-HIRES and VLT-UVES; preferably some Werner lines should be observed for dealing with systematics~\cite{Malec2010}.}
    \end{itemize}
\begin{itemize}
\item{No occurrence of additional DLAs and only a limited amount of metal absorption systems along the same line of sight; the Q$2348-011$ absorber is an example of an unfavorable situation with an additional strong DLA and a few very strong H I absorbers, as well as several metal bearing systems.}
    \end{itemize}
\begin{itemize}
\item{In principle, a firm constraint on the velocity structure of the H$_2$ absorber is only possible with the comprehensive fitting method. The present example of Q$2348-011$ demonstrates the applicability of this method in case of a complex structure with 7 velocity components. For line-by-line fitting methods the H$_2$ spectrum should preferably exhibit a single absorber; that a, e.g. two-component model, is statistically worse than a single component model should be demonstrated.}
\end{itemize}

\noindent
We have inspected the existing literature on high-redshift H$_2$ absorbers in order to identify possibly useful systems that may be targeted in future for $\mu$-variation. In Table~\ref{tab:1} all systems in which H$_2$ has been detected so far are listed. Relevant details for all systems, such as absorption and emission redshifts, their position on the sky, the obtained column densities for H$_2$, HD and H I, and the magnitude are listed as well. Inspection of these systems leads to a quality assessment of the known H$_2$ absorber systems. Some of the systems can be discarded based on their low redshift (Q$0551-366$, Q$2318-111$, Q$0013-004$ and Q$1331+170$), on their low H$_2$ column density (Q$2343+125$ and Q$1337+315$), too low brightness, or combinations thereof.
It should be noted that this assessment is linked to the status of present day technology. Future larger telescopes, or space-based telescopes equipped with high resolution spectrometers ($R > 50\,000$) may alter this view.

\begin{table}[hb]
\caption{Listing of known high redshift H$_2$ absorption systems with some relevant parameters. Bessel $R$ photographic magnitude taken from the SuperCOSMOS Sky Survey~\cite{Hambly2001}.
The five high-quality systems analyzed so far are offset.
The column densities $N$(H$_2$), $N$(HD) and $N$(H I) are given on a $\log_{10}$ scale in [cm$^{-2}$]. }
\label{tab:1}
\begin{tabular}{p{1.7cm}p{0.7cm}p{0.6cm}p{1.6cm}p{1.6cm}p{0.8cm}p{0.8cm}p{0.8cm}p{0.8cm}p{1.2cm}}
\hline\noalign{\smallskip}
Quasars & $z_{abs}$ & $z_{em}$     & RA(J2000)  & Decl.(J2000) & $N$(H$_2$) & $N$(HD) & $N$(H I) & $R_{mag}$ & Ref. \\
\noalign{\smallskip}\svhline\noalign{\smallskip}
{\bf Q0347$-$383} & $3.02$  & $3.21$ & 03:49:43.64 & $-$38:10:30.6 &$14.5$ &        & $20.6$ &$17.48$& \cite{Reinhold2006,King2008} \\
{\bf Q0405$-$443} & $2.59$  & $3.00$ & 04:07:18.08 & $-$44:10:13.9 &$18.2$ &        & $20.9$ & $17.34$  & \cite{Reinhold2006,King2008} \\
{\bf Q0528$-$250} & $2.81$  & $2.81$ & 05:30:07.95 & $-$25:03:29.7 &$18.2$ &  $^a$  & $21.1$ & $17.37$  & \cite{King2008} \\
{\bf J2123$-$005} & $2.06$  & $2.26$ & 21:23:29.46 & $-$00:50:52.9 & $17.6$ & $13.8$ & $19.2$ & $15.83$  & \cite{Malec2010} \\
{\bf Q2348$-$011}$^b$ & $2.42$  & $3.02$ & 23:50:57.87 &  $-$00:52:09.9  & $18.4$ &   & $20.5$  & $18.31$  & \cite{Ledoux2006,Noterdaeme2007}$^c$ \\
\\
{\bf Q0013$-$004} & $1.97$  & $2.09$ & 00:16:02.40 & $-$00:12:25.0 &$18.9$ &        & $20.8$ & $17.89$  & \cite{Petitjean2002} \\
{\bf HE0027$-$184}& $2.42$  & $2.55$ & 00:30:23.62 & $-$18:19:56.0 & $17.3$ &        & $21.7$ & $17.37$  & \cite{Noterdaeme2007b} \\
{\bf Q0551$-$366} & $1.96$  & $2.32$ & 05:52:46.18 & $-$36:37:27.5 & $17.4$ &        & $20.5$ & $17.79$  & \cite{Ledoux2002}  \\
{\bf Q0642$-$506} & $2.66$  & $3.09$ & 06:43:26.99 & $-$50:41:12.7 & $18.4$ &        & $21.0$ & $18.06$  & \cite{Noterdaeme2008b} \\
{\bf FJ0812$+$320}& $2.63$  & $2.70$ & 08:12:40.68  & $+$32:08:08.6 & $19.9$ & $15.4$ & $21.4$ & $17.88$ & \cite{Tumlinson2010,Balashev2010} \\
{\bf Q0841$+$129} & $2.37$  & $2.48$ & 08:44:24.24 & $+$12:45:46.5 & $14.5$ &        & $20.6$ & $17.64$   & \cite{Petitjean2000} \\
{\bf Q1232$+$082} & $2.34$  & $2.57$ & 12:34:37.58 & $+$07:58:43.6 & $19.7$ & $15.5$ & $20.9$ & $18.40$  & \cite{Varshalovich2001,Ivanchik2010} \\
{\bf J1237$+$064} & $2.69$  & $2.78$ & 12:37:14.60 & $+$06:47:59.5 & $19.2^d$& $14.5$& $20.0$ &  $18.21$ & \cite{Noterdaeme2010} \\
{\bf Q1331$+$170}$^e$ & $1.78$  & $1.78$ & 13:33:35.81 & $+$16:49:03.7 & $19.7$ & $14.8$ & $21.2$ & $16.26$  & \cite{Balashev2010,Cui2005} \\
{\bf Q1337$+$315} & $3.17$  & $3.17$ & 13:37:24.69 & $+$31:52:54.6 & $14.1$ &        & $21.4$ & $ 18.08  $   & \cite{Srianand2010} \\
{\bf Q1439$+$113} & $2.42$  & $2.58$ & 14:39:12.04 & $+$11:17:40.5 & $19.4$ & $14.9$ & $20.1$ & $18.07$  & \cite{Noterdaeme2008} \\
{\bf Q1443$+$272} & $4.22$  & $4.42$ & 14:43:31.18 & $+$27:24:36.4 & $18.3$ &        & $21.0$ & $18.81$  & \cite{Ledoux2006b} \\
{\bf Q1444$+$014} & $2.08$  & $2.21$ & 14:46:53.04 & $+$01:13:56.0 & $18.3$ &        & $20.1$ & $18.10$  & \cite{Ledoux2003} \\
{\bf Q2318$-$111} & $1.99$  & $2.56$ & 23:21:28.69 & $-$10:51:22.5 & $15.5$ &        & $20.7$ & $17.67$  & \cite{Noterdaeme2007b} \\
{\bf Q2343$+$125} & $2.43$  & $2.52$ & 23:46:25.42 & $+$12:47:43.9 & $13.7$ &        & $20.4$ & $20.18$  & \cite{Petitjean2006,Dessauges2004} \\
\noalign{\smallskip}\hline\noalign{\smallskip}
\end{tabular}
$^a$ A recent re-analysis has shown that this system contains HD~\cite{King-fut}.\\
$^b$ Multiple DLAs in this system.\\
$^c$ Studied in this work for $\mu$-variation.\\
$^d$ The strongest H2 component does not coincide with the centre of the HI absorption.\\
$^e$ Observed with the Hubble Space Telescope.\\
\end{table}

Based on the considerations given we have identified five quasar systems that have the potential to achieve a constraint at the $\delta(\Delta\mu/\mu) \leq 5 \times 10^{-6}$ level. These systems have been investigated to some extent, but larger numbers of exposures yielding better S/N, and ThAr attached calibration frames will be needed to reach the desired accuracy. Of the five systems, four have been observed with VLT-UVES, while one (FJ$0812+320$) has been studied with Keck-HIRES.
\begin{itemize}
\item{{\bf Q0642$-$506} with redshift $z_{abs}=2.66$ provides a good number of H$_2$ lines with a high column density giving rise to saturation of only the strongest lines. The occurrence of only a single velocity components renders the spectrum relatively simple, also accessible for a line-by-line treatment. Spectra are shown in \cite{Noterdaeme2008b}.}
    \end{itemize}
\begin{itemize}
\item{{\bf HE0027$-$184} is of similar good quality. It has a favorable magnitude and a molecular column density which allows for observation of a large number of lines to be observed. Indeed in the archived spectra with data observed in 2004 many lines are found, possibly even up to $J=6$ \cite{Noterdaeme2007b}.}
    \end{itemize}
\begin{itemize}
\item{{\bf FJ0812$+$320} has a strong saturated H$_2$ component, which makes it less ideal. The system has been studied in particular for its abundant presence of the widest variety of metal lines~\cite{Prochaska2003} and of HD \cite{Tumlinson2010,Balashev2010}. The system is nevertheless of importance because it is the only system of reasonable quality at northern declination, except for the special $z=4.22$ system discussed below.}
    \end{itemize}
\begin{itemize}
\item{{\bf J1237$+$064} at $z_{abs}=2.69$ shares some of the disadvantages of the previous system, in particular the very high column density giving rise to  saturation. However, the velocity structure exhibits two weaker components, such that high-$J$ lines can be obtained in the strong components, and the low-$J$ lines in the weak components; HD is abundantly present and can be included in the analysis. This makes J1237$+$064 a target with a high potential. Further details on the system in \cite{Noterdaeme2010}.}
    \end{itemize}
\begin{itemize}
\item{{\bf Q1443$+$272} is a special case, and its unusually high redshift ($z_{abs}=4.22$) alters some of the arguments. The system has a very low brightness, but that is partly compensated because the H$_2$ lines shift further into the visible, where scattering is less and CCDs are more sensitive. In addition the high redshift shifts many more H$_2$ lines into the observable range. However, for these high redshift objects the Lyman-$\alpha$ forest becomes denser, which is a disadvantage. In summary this system has a good potential, if sufficient observation time is made available. Further details on the system in \cite{Ledoux2006b}.}
    \end{itemize}

\section{Conclusion}
\label{sec:4}

We have reviewed the status of investigations into variation of the proton-electron mass ratio from analyses of highly redshifted H$_2$ absorption systems. As of now five results have been obtained putting a tight constraint at the $\Delta\mu/\mu < 1 \times 10^{-5}$ level. We present spectroscopic data on the fifth H$_2$ absorption system, Q$2348-011$, and have shown how a complex velocity structure can be unraveled using the comprehensive fitting method. An attempt is made to list in full all relevant quasar systems where H$_2$ has been observed so far, followed by an assessment as to which systems could produce a competitive result on $\Delta\mu/\mu$, provided that high-quality well-calibrated observations are performed.

\begin{acknowledgement}
This work is based on observations carried out at the European Southern
Observatory (ESO) under program ID 79.A-0404 (PI Ubachs), with the
UVES spectrograph installed at the VLT Kueyen UT2 on Cerro Paranal, Chile.
The authors wish to thank John Webb, Julian King and Steve Curran (Sydney),
Adrian Malec (Melbourne) and Freek van Weerdenburg (Amsterdam) for fruitful
discussions.
\end{acknowledgement}

%
%

\begin{thebibliography}{99.}%
\bibitem{Thompson1975} R. Thompson,  Astroph. Lett. {\bf 16}, 3 (1975).
\bibitem{Ubachs2007} W. Ubachs. R. Buning, K.S.E. Eikema, E. Reinhold,
 J. Mol. Spectr. {\bf 241}, 155 (2007).
\bibitem{Hinnen1994} P.C. Hinnen, W. Hogervorst, S. Stolte, W. Ubachs,
 Can. J. Phys. {\bf 72}, 1032 (1994).
\bibitem{Philip2004} J. Philip, J.P. Sprengers, Th. Pielage, \emph{et al.},
    Can. J. Chem. {\bf 82}, 713 (2004).
\bibitem{Ubachs2004} W. Ubachs, E. Reinhold,
 Phys. Rev. Lett. {\bf 92}, 101302 (2004).
\bibitem{Hollenstein2006} U. Hollenstein, E. Reinhold, C. A. de Lange, W. Ubachs,
  J. Phys. B {\bf 39}, L195 (2006).
\bibitem{Reinhold2006} E. Reinhold, R. Buning, U. Hollenstein, \emph{et al.},
 Phys. Rev. Lett.  {\bf 96}, 151101 (2006).
\bibitem{Ivanov2008a} T.I. Ivanov, M.O. Vieitez, C.A. de Lange, W. Ubachs,
 J. Phys. B. {\bf 41}, 035702 (2008).
\bibitem{Hinnen1995} P.C. Hinnen, S.E. Werners, W. Hogervorst, S. Stolte, \emph{et al.},
 Phys. Rev. A {\bf 52}, 4425 (1995).
\bibitem{Ivanov2008b} T.I. Ivanov, M. Roudjane, M.O. Vieitez, \emph{et al.},
 Phys. Rev. Lett. {\bf 100}, 093007 (2008).
\bibitem{Ivanov2010} T.I. Ivanov, G.D. Dickenson, M. Roudjane, \emph{et al.},
 Mol. Phys. {\bf 108}, 771 (2010).
\bibitem{Hannemann2006} S. Hannemann, E.J. Salumbides, S. Witte, \emph{et al.},
 Phys. Rev. A {\bf 74}, 062514 (2006).
\bibitem{Salumbides2008} E.J. Salumbides, D. Bailly, A. Khramov, \emph{et al.},
 Phys. Rev. Lett.  {\bf 101}, 223001 (2008).
\bibitem{Bailly2010} D. Bailly, E.J. Salumbides, M. Vervloet, W. Ubachs,
 Mol. Phys. {\bf 108}, 827 (2010).
\bibitem{Malec2010} A.L. Malec, R. Buning, M.T. Murphy, \emph{et al.},
 MNRAS {\bf 403}, 1541 (2010).%
\bibitem{Abgrall1994} H. Abgrall, E. Roueff, F. Launay, J.-Y. Roncin,
 Can. J. Phys. 72, 856 (1994).
\bibitem{Abgrall2000} H. Abgrall, E. Roueff, I. Drira,
 Astron. Astrophys. Suppl. 141, 297 (2000).
\bibitem{VPFIT} VPFIT software developed by R.F. Carswell; see http://www.ast.cam.ac.uk/$\sim$rfc/vpfit.html
\bibitem{King2008} J.A. King, J.K. Webb, M.T. Murphy, R.F. Carswell,
 Phys. Rev. Lett. {\bf 101}, 251304 (2008).
\bibitem{Murphy2008} M.T. Murphy, V.V. Flambaum, S. Muller, C. Henkel,
 Science {\bf 320}, 1611 (2008).%
\bibitem{Henkel2009} C. Henkel, K.M. Menten, M.T. Murphy, N. Jethava, \emph{et al.},
 Astron. Astroph. {\bf 500}, 725 (2009).%
\bibitem{Petitjean2006} P. Petitjean, C. Ledoux, P. Noterdaeme, R. Srianand,
 Astron. Astroph. {\bf 456}, L9 (2006).
\bibitem{Ledoux2006} C. Ledoux, P. Petitjean, J.P.U. Fynbo, P. Moller, R. Srianand,
 Astron. Astroph. {\bf 457}, 71 (2006).
\bibitem{Noterdaeme2007} P. Noterdaeme, P. Petitjean, R. Srianand, C. Ledoux, \emph{et al.},
 Astron. Astroph. {\bf 469}, 425 (2007).%
\bibitem{Murphy2007} M.T. Murphy, P. Tzanavaris, J.K. Webb, C. Lovis,
 MNRAS {\bf 378}, 221 (2007).%
\bibitem{Webb2010} J.K. Webb, J.A. King, M.T. Murphy, V.V. Flambaum, \emph{et al.},
 arXiv:1008.3907.%
\bibitem{JB2011} J. Bagdonaite, M.T. Murphy, R. Buning, L. Kaper, W. Ubachs, to be published.%
\bibitem{Wendt2008} M. Wendt, D. Reimers,
 Eur. J. Phys. D Spec. Top. {\bf 163}, 197 (2008).
\bibitem{Thompson2009} R.I. Thompson, J. Bechtold, J.H. Black, D. Eisenstein, \emph{et al.},
 Astroph. J. {\bf 703}, 1648 (2009).
\bibitem{Wendt2010} M. Wendt, P. Molaro,
 Astron. Astroph. {\bf 526}, A96 (2010).
\bibitem{Srianand2010} R. Srianand, N. Gupta, P. Petitjean, P. Noterdaeme, C. Ledoux,
 MNRAS {\bf 405}, 1888 (2010).
\bibitem{Veron2010} M.-P. V\'{e}ron-Cetty and P. V\'{e}ron,
 Astron. Astroph. {\bf 518}, A10 (2010).
\bibitem{Noterdaeme2008b} P. Noterdaeme, C. Ledoux, P. Petitjean, R. Srianand,
 Astron. Astroph. {\bf 481}, 327 (2008).
\bibitem{Noterdaeme2007b} P. Noterdaeme, C. Ledoux, P. Petitjean, F. Le Petit, \emph{et al.},
 Astron. Astroph. {\bf 474}, 393 (2007).
\bibitem{Prochaska2003} J.X. Prochaska, J.C. Howk, A.M. Wolfe, Nature {\bf 423}, 57 (2003).
\bibitem{Tumlinson2010} J. Tumlinson, A.L. Malec, R.F. Carswell, \emph{et al.},
 Astroph. J. Lett. {\bf 718}, L156 (2010).
\bibitem{Balashev2010} S.A. Balashev, A.V. Ivanchik, D.A. Varshalovich,
 Astron. Lett. {\bf 36}, 761 (2010).
\bibitem{Noterdaeme2010} P. Noterdaeme, P. Petitjean, C. Ledoux, S. Lopez, \emph{et al.}, Astron. Astroph. {\bf 523}, A80 (2010).
\bibitem{Ledoux2006b} C. Ledoux, P. Petitjean, R. Srianand,
 Astroph. J. Lett. {\bf 640}, L25 (2006).
\bibitem{Hambly2001} N.C. Hambly, H.T. MacGillivray, M.A. Read, \emph{et al.},
 MNRAS {\bf 326}, 1279 (2001).
\bibitem{Petitjean2002} P. Petitjean, R. Srianand, C. Ledoux, MNRAS {\bf 332}, 383 (2002).
\bibitem{Ledoux2002} C. Ledoux, R. Srianand, P. Petitjean,
 Astron. Astroph. {\bf 392}, 781 (2002).
\bibitem{Petitjean2000} P. Petitjean, R. Srianand, C. Ledoux,
 Astron Astroph. {\bf 364}, L26 (2000).
\bibitem{Varshalovich2001} D.A. Varshalovich, A.V. Ivanchik, P. Petitjean, \emph{et al.},
 Astron. Lett. {\bf 27}, 683 (2001).
\bibitem{Ivanchik2010} A.V. Ivanchik, P. Petitjean, S.A. Balashev, R. Srianand, \emph{et al.},
 MNRAS {\bf 404}, 1583 (2010).
\bibitem{Cui2005} J. Cui, J. Bechtold, J. Ge, D.M. Meyer, Astroph. J. {\bf 633}, 649 (2005).
\bibitem{Noterdaeme2008} P. Noterdaeme, P. Petitjean, C. Ledoux, \emph{et al.},
 Astron. Astroph. {\bf 491}, 397 (2008).
\bibitem{Ledoux2003} C. Ledoux, P. Petitjean, R. Srianand,
 MNRAS {\bf 346}, 209 (2003).
\bibitem{Dessauges2004} M. Dessauges-Zavadsky, F. Calura, J.X. Prochaska, \emph{et al.},
 Astron. Astroph. {\bf 416}, 79 (2004).
\bibitem{King-fut} J.A. King, W. Ubachs, M.T. Murphy, J.K. Webb, L. Kaper,
 to be published.
\end{thebibliography}
%

\end{document}